# *Ab initio* Calculations of the Vibrational Modes of MnAs and $Ga_{1-x}Mn_xAs$

## H. W. Leite Alves

*Departamento de Ciências Naturais, UFSJ, C. P.: 110, CEP: 36.300-000 São João del Rei, MG, Brazil*

**Abstract.** In this work, we present our theoretical results for the equation of state and the phonon dispersions of MnAs, as well as the Mn concentration dependence of both the lattice parameter and the phonon frequencies of the cubic GaMnAs alloys. The results are in good agreement with the experimental results whenever this comparison is possible. Based on the obtained results, the lattice constants and the phonon frequencies of the alloys do not obey the Vegard rule.

The introduction of magnetic or spin-related functions into semiconductor materials, such as MnAs and its alloys, is a new challenge in materials science that has been increased in the last years [1]. Specifically, in the semiconductor spintronics, one of the important issues is the development of ferromagnet/semiconductor hybrid structures, in which the spin degree of freedom can be used. Despite the fact that epitaxial techniques offer new opportunities for exploration in this direction, very little is known about this hybrid structures. Moreover, the experimental and theoretical data on the phonon modes available in the literature are very scarce. In this work, by using the Density Functional Theory (DFT) within both the Local Density Approximation (LDA) and Generalized Gradient Approximation (GGA) for the exchange-correlation term, plane-wave description of the wavefunctions and the pseudopotential method (abinit code [2]), we have calculated the the equation of state and the vibrational modes for the MnAs and the $Ga_{1-x}Mn_xAs$ alloys. We have used the Troullier-Martins pseudopotentials, and the phonons were obtained by means of the Density-Functional Perturbation Theory. Moreover, in order to describe the alloys, we have used the Average Crystal Approximation, as defined by Ghosez *et al* [3].

Before to proceed the phonon calculations, we have minimized the total energy of the MnAs, in both zincblende and NiAs structures, as a function of the lattice parameters and the internal atomic positions. For the the hexagonal modification, we have obtained $a_0$ = 3.84 (3.50) Å, $c/a$ = 1.48 (1.48), $B_0$ = 0.72 (1.77) Mbar, and $B_0'$ = 3.9 (4.2) for the GGA (LDA) calculations. For the zincblende structure, we found $a_0$ = 5.87 (5.23) Å, $B_0$ = 0.44 (1.16) Mbar, and $B_0'$ = 3.8 (4.1), respectively. Our results are in good agreement with the available theoretical and experimental results [4]. Also, the results show that the NiAs is more stable than the zincblende one.

In Figure 1, we display our calculated LDA phonon dispersion as well as the phonon density of states for the hexagonal MnAs, which the $B_{1u}$ mode (silent mode) has the highest frequency, 298.6 cm$^{-1}$. The calculated GGA phonon dispersion shows the same trend but, with lower frequencies when compared with the LDA ones, due to the lower value of its bulk modulus. As an example, the obtained GGA frequency for the $B_{1u}$ mode is 266.6 cm$^{-1}$. Moreover, another remarkable feature between the two calculations was the calculated frequency split of the $E_{2u}$ and $A_{2u}$ modes. The obtained values, 3.6 cm$^{-1}$ (in the LDA calculations) and 23.5 cm$^{-1}$ (in the GGA ones), show this difference.

In Table 1, we show the calculated LDA and GGA phonon frequencies for both hexagonal and cubic MnAs. We would like to mention that the values for the LO($\Gamma$) and TO($\Gamma$) vibrational modes in the cubic modification are higher than the GaAs LO($\Gamma$) one, 291 cm$^{-1}$. Unfortunately, there is no experimental data reported in the literature in order to compare with our obtained results.

Our calculated LDA lattice parameter dependence on the Mn composition $x$ for the $Ga_{1-x}Mn_xAs$ alloys in the diluted regime (x < 6 %) is shown in Figure 2. Only at this region this dependence is linear, and our results were fitted by a straight line, $(0.77\pm0.02).x + (5.626\pm0.001)$, which is in good agreement with the experimental results of Ohno [5]. When $x > 10\%$, this dependence has a negative bowing, reflecting strong changes in the alloy structure. We can infer, from this feature, that there is a possibility of segregation of this alloy at this range of Mn concentration, as observed in the InGaN ones [6], once that these atoms are fast diffusers in GaAs, and that they have the ability of make impurity clusters in this crystal.

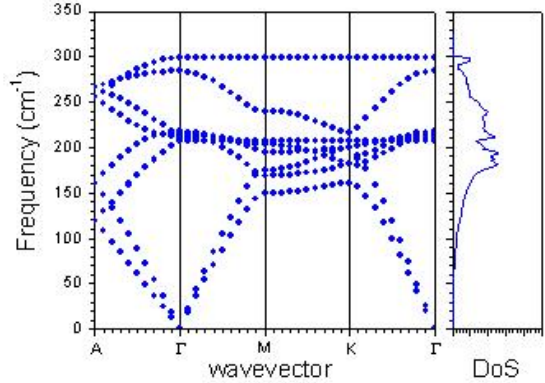

**FIGURE 1.** Calculated LDA phonon dispersion and density of states for the hexagonal MnAs.

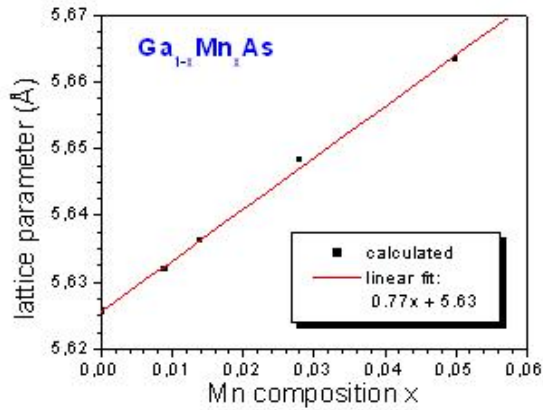

**FIGURE 2.** Calculated Mn composition $x$ dependence of the lattice parameter for the cubic GaMnAs alloy (diluted regime).

In Figure 3, we display our calculated LDA frequency shifts ($\Delta\omega$) dependence with the Mn composition $x$ for the LO and TO modes of the $Ga_{1-x}Mn_xAs$ alloys at the $\Gamma$ point of the Brillouin zone in the diluted regime. As shown in this Figure, the results were fitted by polynomials of the second degree, one for each mode. As a consequence, there is a positive bowing in the Mn concentration dependence of phonon frequencies of both LO and TO modes, indicating that, for the phonon frequencies, this material does not obey the Vegard rule, as observed in the Raman experiments [7].

In summary, we have presented our obtained results for the phonon frequencies of the MnAs and the GaMnAs alloy in the diluted regime. A complete description of our obtained results will be published elsewhere.

**TABLE 1.** Calculated phonon modes for MnAs.

| Mode | LDA | GGA |
|---|---|---|
| Hexagonal structure | | |
| $E_{2u}$ | 207.8 | 167.9 |
| $A_{2u}$ | 211.4 | 191.4 |
| $E_{2g}$ | 213.5 | 192.3 |
| $B_{2g}$ | 217.5 | 192.4 |
| $E_{1u}$ | 284.0 | 254.5 |
| $B_{1u}$ | 298.6 | 266.6 |
| Cubic structure | | |
| TO($\Gamma$) | 319.0 | 247.5 |
| LO($\Gamma$) | 339.7 | 309.0 |

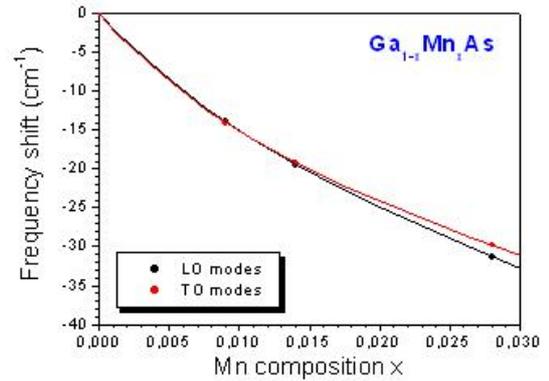

**FIGURE 3.** Calculated Mn composition $x$ dependence of the LO($\Gamma$) and TO($\Gamma$) frequency shifts for the cubic GaMnAs alloy (diluted regime).

## ACKNOWLEDGMENTS

This work was supported by the FAPEMIG project CEX 80953/04, Brazil.